\documentclass[epj]{svjour}
\usepackage{graphics}
\usepackage{epsfig,array}
\usepackage{color}
\newcommand{\be}{\begin{eqnarray}}
\newcommand{\ee}{\end{eqnarray}}

\newcommand{\er}{$\pm$}
\begin{document}

\title{The study of the proton-proton collisions at the beam momentum 1628 MeV/c}

\titlerunning{The study of the proton-proton collisions...}

\author{K.N.~Ermakov      \and
        V.I.~Medvedev     \and
        V.A.~Nikonov      \and
        O.V.~Rogachevsky  \and
        A.V.~Sarantsev    \and
        V.V.~Sarantsev    \and
        S.G.~Sherman
}

\authorrunning{K.N.~Ermakov \it et al.}

\institute{Petersburg Nuclear Physics Institute, Gatchina 188300, Russia}

\date{Received: \today / Revised version:}

\abstract {The detailed investigation of the single pion production
reactions $pp\rightarrow pn \pi^{+}$ and $pp \rightarrow pp \pi^{o}$
at the incident proton momentum 1628 MeV/c has been carried out. The
data are analyzed in the framework of the event-by-event maximum
likelihood method together with the $pp\rightarrow pp \pi^{0}$ data
measured earlier in the energy region below 1 GeV. At 1628 MeV/c the
largest contributions stem from the $^{3}P_{2}$, $^{3}P_{1}$,
$^{3}P_{0}$, $^{1}D_{2}$ and $^{3}F_{2}$  initial partial waves.}


\maketitle

\section{Introduction}
\label{intro}
Nucleon-nucleon interaction is one of the most important processes
in nuclear and particle physics and extensively has been studied
over a wide energy range. The single pion production in the $NN$
interactions is the main inelastic process at energies below 1 GeV.
Despite the fact that a lot of experiments have been performed, many
questions on this process are not yet answered. One of them is the
question about a contribution of the isoscalar ($I=0$) channel to
the inelastic neutron-proton collisions. Since the neutron-proton
scattering amplitude contains both isoscalar and isovector ($I=1$)
parts, a detailed investigation of the single pion production in the
$pp$ collisions (isovector contribution only) might give the most
accurate information about the isovector channel which, in
combination with the neutron-proton data, would allow one to extract
correctly the contribution of the isoscalar channel.

     Various theoretical models, more or less successful, were put forward
while the data on the pion production in the $NN$ collisions were
accumulated. Most of them are constructed for the energy region near
the production threshold and can not be applied at higher energies.
For the energy range about 1 GeV the one-pion exchange (OPE) model
\cite{1} assumes a dominance of the one-pion exchange contribution
to the inelastic amplitude. Pole diagram matrix elements were
calculated using a beforehand form factor. The form factor function
was obtained by fitting experimental data, so in fact this was a
semiphenomenological model. In addition, only the {\it P}$_{33}$
partial wave was taken into account in the intermediate {\it $\pi$N}
channel \cite{2}. Nevertheless, this model predicts (up to a
normalization factors) with a reasonable accuracy the differential
spectra of the $pp\rightarrow pn\pi^+$ and $pp\rightarrow pp\pi^0$
reactions in the energy range 600-1300 MeV \cite{2,3}. At the same
time discrepancies between the measured total cross sections for
these reactions and the model predictions are sizeable enough.

 It should be noted that the experimental data on the differential spectra of
{\it pp $\rightarrow$ pp$\pi^0$} reaction near the energy of 1 Gev
are more scarce than the data for {\it pp $\rightarrow$ pn$\pi^+$}
channel. The KEK data \cite{Shimizu:1982dx} contain the information
on total cross sections only. Our data on the differential spectra
of the {\it pp $\rightarrow$ pp$\pi^0$} reaction at the energies 990
MeV and 900 MeV were published earlier \cite{5}. It would be
important to perform an accurate measurement of the differential
cross sections in the middle of this energy region. Here we present
our investigation of the {\it pp $\rightarrow$ pn$\pi^+$} and {\it
pp $\rightarrow$ pp$\pi^0$} reactions at 940 MeV and determine
contributions from various partial waves to the single pion
production processes.

\section{Experiment}
\label{sec:1}

 The experiment was performed at PNPI 1 GeV synchrocyclotron. The events
were registered by a 35 cm hydrogen bubble chamber disposed in the
1.48 T magnetic field. The proton beam (after corresponding degrader
for the momentum 1628 MeV/c) was formed by three bending magnets and
by eight quadrupole lenses. The incident proton momentum value was
inspected by the kinematics of the elastic scattering events. The
accuracy of the incident momentum value and momentum spread was
about 0.5 MeV/c and 20 MeV/c (FWHM) correspondingly. A total of
$4.6\times10^5$ stereoframes were obtained. The frames were double
scanned to search for events due to an interaction of the incident
beam. The double scanning efficiency was determined to be 99.95\%.
Approximately 8$\times10^3$ two-prong events were used for
subsequent analysis.

 The 2-prong events selected in the fiducial volume of the chamber were
measured and geometrically reconstruc-
ted. The reconstructed events were
kinematically fitted to the following reaction hypotheses:
\be
p+p&\to& p+p,       \\
p+p&\to& p+n+\pi^+, \\
p+p&\to& p+p+\pi^0, \\
p+p&\to& d+\pi^+,   \\
p+p&\to& d+\pi^{+}+\pi^0.
\ee

 The identification of the events was performed on the $\chi^2$
criteria with confidence level less then 1$\%$. If the event had a
good $\chi^2$ for the elastic kinematic (4C-fit), it was considered
as elastic one. Stretch functions for the three kinematical
variables of a track (the inverse of the momentum, the azimuthal and
dip angles) were examined for elastic scattering events to make sure
that their errors and the error for bubble chamber magnetic field
were properly given. If there was only one acceptable fit for the
event it was identified as belonging to this hypothesis (with a
check of stopping $\pi^{+}$ track on the presence of the $\pi
\rightarrow \mu \rightarrow e$ decay). If several inelastic versions
revealed a good $\chi^2$, we used visual estimation of the bubble
density of the track to distinguish between proton (deuteron) and
pion. For few events, even after repeated measurements, the fit
revealed only one acceptable hypothesis but with a large $\chi^2$.
If such event had $\chi^2$ less than 50 for the 4-constraint fit or
less than 20 for the 1-constraint fit, it was taken into account for
the calculation of the cross section value of the process
corresponding to this hypothesis but not included in differential
spectra.

There were events which failed to fit any hypothesis. These no-fit
events were investigated on the scanning table, and most of them
appeared to be events with a secondary track undergoing one more
scattering near a primary vertex.

There were also events unfit for the measurements, e.g. events with
a bad vertex or superimposed tracks. The number of such events was
counted approximately to be 7\%. The total number of 2-prong events
which did not pass the measurement and fitting procedures was
counted to be less than 10\%. These unidentified events were
apportioned to the fraction of the fitted hypotheses for the total
cross section measurements.

Missing mass distributions for accepted events show clear peaks - at
zero for the elastic scattering, at the $\pi^0$ mass squared for
single neutral pion production and at the neutron mass squared for
the process with neutron in the final state.

The standard bubble chamber procedure was used to obtain absolute
cross sections \cite{3} for the elastic and single pion production
reactions. These values together with statistic of the events are
listed in Table~\ref{tot_cs}.

\begin{table}
\begin{center}
\caption{\label{tot_cs} Numbers of events and cross section values
at the beam momentum 1628 MeV/c.}

\begin{tabular}{|c|cccc|}

\hline
pp$\to$ & pp &   pn$\pi^+$ & pp$\pi^0$ &  d$\pi^+$ \\
\hline
events &3442  &   3014  &   696  &    80 \\
\hline
$\sigma$  mb & 21.2$\pm $0.7 & 17.6$\pm $0.6 &  4.48$\pm $0.20 &  0.47$\pm $0.05 \\
\hline
\end{tabular}
\end{center}
\end{table}

\section{Experimental results and discussion}
\label{sec:3}
\subsection{Elastic scattering}

\begin{figure}[ht]
\centerline{\epsfig{file=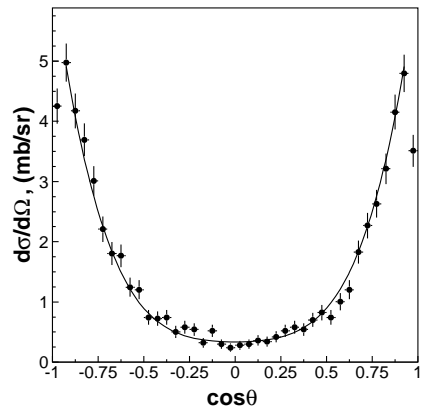,width=0.30\textwidth}}
\caption{\label{elastic} Elastic differential cross section. The
curve is result of the Legender polynomial fit of the -0.95$\leq
cos\theta \leq$0.95 angular range.}
\end{figure}

The measured differential cross section of the elastic pp-scattering
in c.m.s. of the reaction is shown in Fig.~\ref{elastic}. The
elastic cross section value given in Table~\ref{tot_cs} was
calculated as $4\pi A_0$ where $A_0$ is the coefficient in front of
the Legender polynomial $P_0(z)$ in fitting the angular distribution
by the Legender expansion in the angular range -0.95$\leq cos\theta
\leq$0.95.

This range was chosen in order to take into account a loss of
events due to forward elastic scattering when a slow proton had
a short recoil path and could not be seen in the bubble chamber (at the
momentum less than 80 MeV/c) or was missed under the scanning. This
interval was determined by examining the stability of the Legender
coefficients with decreasing fitted range of the angular
distribution. Our value of the total elastic cross section is less
than that given in works \cite{Shimizu:1982dx,6} approximately by 13$\%$. We do not
know the reason for this discrepancy: it could arise due to i) an
insufficient control of the loss of events with small scattering
angles or ii) an uncorrect calculation of the millibarn-equivalent
in the chamber. To control the loss of events in the
forward scattering we use the short-cut range for fitting of the
differential cross section which should repair this shortcoming. Of course we
have about 7$\%$ unmeasured events which were apportioned to the
fraction of the fitted hypotheses for the cross section
calculations. But it is difficult to believe that the fraction of
missing elastic events is much larger. The missing events have
superimposed second tracks: this topology is not the elastic
scattering one, except for events with disposition of the scattering
plane strongly in the normal direction to the frame of a film.
As concerns the millibarn-equivalent calculation, we would like to pay attention that
the values for the single pion production cross sections given in
Table~\ref{tot_cs} are in a fairly good agreement with the above mentioned work
\cite{Shimizu:1982dx}.

\subsection{ The single pion production reactions. Comparison with OPE model}

 The OPE model \cite{1,2} describes the single pion production reactions by
four pole diagrams with the $\pi^0$ or $\pi^+$ exchanges. The main
evidence for pole diagram contributions would be an observation
of a peak in the momentum transfer distribution from the target
particle to the secondary proton in the $pp \rightarrow pp\pi^0$
process or, for example, to the secondary neutron in the
$pp \rightarrow pn\pi^+$ process. Since there is no
difference between final protons in the $pp \rightarrow
pp\pi^0$ reaction, it is difficult to separate the contribution
from a certain diagram experimentally.

\begin{figure}[ht]
\centerline{\epsfig{file=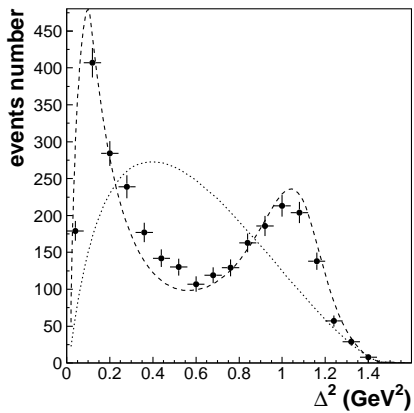,width=0.30\textwidth}}
\caption{\label{d2pip} Four-momentum transfer $\Delta^2$
distribution for the $pp \rightarrow pn\pi^+$ reaction. The dashed
curve is the OPE calculation and dotted one shows shape of the phase
space.}
\end{figure}

\begin{figure}[ht]
\centerline{\epsfig{file=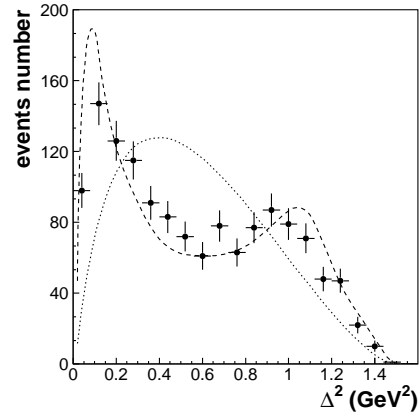,width=0.30\textwidth}}
\caption{\label{d2pi0} Four-momentum transfer $\Delta^2$
distribution for the $pp \rightarrow pp\pi^0$ reaction. The dashed
curve is the OPE calculation and dotted one shows shape of the phase
space.}
\end{figure}

 Figures~\ref{d2pip}, \ref{d2pi0} show the square momentum transfer
$\Delta^2 =-(p_t-p_f)^2$ distribution for the $pp\rightarrow
pn\pi^+$ and $pp \rightarrow pp\pi^0$ reactions, where $p_t$ and
$p_f$ are four-momenta of the target proton and one of the final
nucleons correspondingly. The OPE model calculations, normalized to
the total number of experimental events, are shown there as dashed
lines and phase space distribution as dotted ones.


One can see that the OPE model describes qualitatively well the
$\Delta^2$ distribution for both reactions studied. It is remarkable,
because only the $P_{33}$ wave is
taken into account in the intermediate $\pi N$ scattering. It
could be that this distribution is mainly sensitive
to the pole diagram propagator and a more complicated structure
of the amplitude manifests itself in other distributions.

\begin{figure}[ht]
\centerline{\epsfig{file=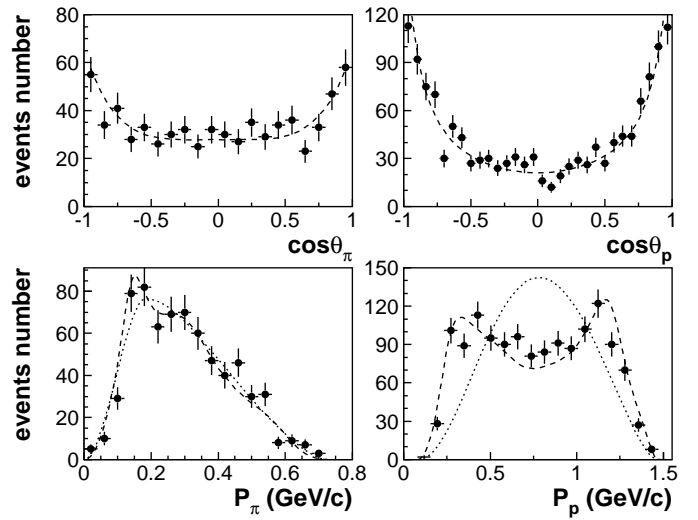,width=0.49\textwidth}}
\caption{\label{opepi0} C.m.s. angular distributions
and laboratory momentum spectra of the final particles of the $pp\to
pp\pi^0$ reaction. The dashed curves are predictions from the OPE model,
the dotted curves represent the phase space.}
\end{figure}
\begin{figure}[ht]
\centerline{\epsfig{file=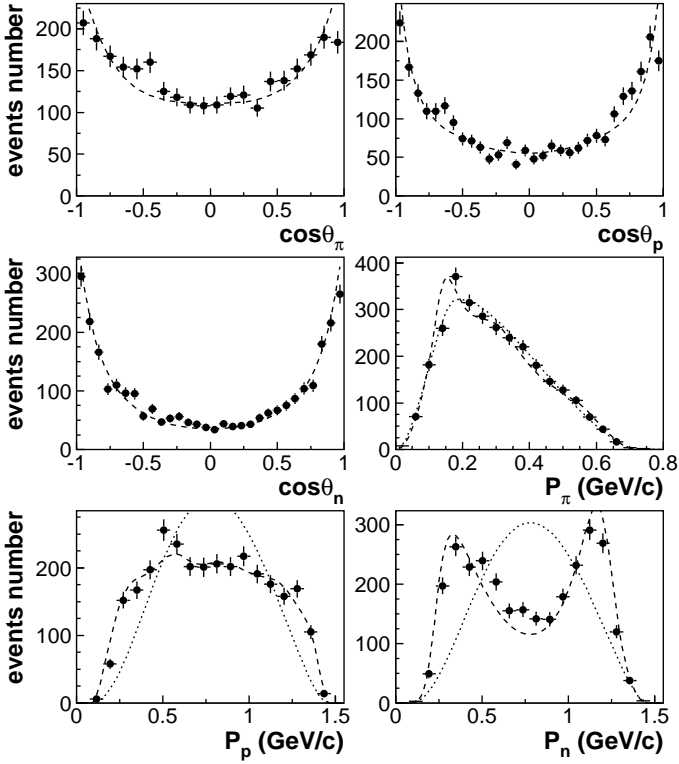,width=0.49\textwidth}}
\caption{\label{opepip} C.m.s. angular distributions
and laboratory momentum spectra of the final particles of the $pp\to
pn\pi^+$ reaction. The dashed curves are predictions from the OPE model,
the dotted curves represent the phase space.}
\end{figure}

 Figs.~\ref{opepi0} and \ref{opepip} present angular distributions of
the final particles in the
c.m.s. of the reaction as well as particle momentum distributions in
the laboratory frame for the $pp \rightarrow pp\pi^0$ and
$pp\rightarrow pn\pi^+$ reactions correspondingly. One can
see again that the agreement of the OPE calculations with the experimental
data is fairly good. The qualitative
agreement is also observed for other spectra, except for angle
distributions of the final particles in the helicity system.

 As noted earlier in \cite{3}, although the OPE model provides a qualitative
description of most differential spectra, it disagrees with the
total cross section values for the $pp \rightarrow pp\pi^0$ and
$pp\rightarrow pn\pi^+$ reactions. It means that taking into account
the P$_{33}$ ($\Delta(1232)$ isobar) intermediate state only is not
enough for an adequate description of these reactions and a
comprehensive partial wave analysis is needed.

\section{Partial wave analysis}

To extract contributions from different partial waves we apply
an event-by-event partial wave analysis (PWA) ba\-sed on the maximum
likelihood method.

For the production of three particles with the 4-mo\-menta
$q_i$ from two particles colliding with 4-momenta $k_1$ and
$k_2$, the cross section  is given by:
\be
d\sigma=\frac{(2\pi)^4|A|^2}{4|\vec k|\sqrt{s}}\,
d\Phi_3(P,q_1,q_2,q_3)\;, \qquad P\!=\!k_1\!+\!k_2\;,
\ee
where $A$ is the reaction amplitude, $\vec k$ is the 3-momentum of
the initial particle calculated in the c.m.s. of the reaction,
$s=P^2=(k_1+k_2)^2$ and $d\Phi_3$ is the invariant three-particle
phase volume.

The total amplitude can be written as a sum of partial wave
amplitudes as follows:
\be
A=\sum\limits_\alpha A^\alpha_{tr}(s) Q^{in}_{\mu_1\ldots\mu_J}(S
LJ)A_{2b}(i,S_2L_2 J_2)(s_i)\times
\nonumber\\
Q^{fin}_{\mu_1\ldots\mu_J}(i,S_2L_2J_2S'L'J)\ .
\ee
Here $S,L,J$ are spin, orbital momentum and total angular momentum
of the $pp$ system, $S_2,L_2,J_2$ are spin, orbital momentum and
total angular momentum of the two-particle system in the final
state and $S',L'$
are spin and orbital momentum between two-particle system and the
third particle with momentum $q_i$. The invariant mass of two-body
system can be calculated as $s_i=(P-q_i)^2$. The multiindex
$\alpha$ denotes all possible combinations of the
$S,L,J,S_2,L_2,J_2,S',L'$ and $i$,  $A^\alpha_{tr}(s)$ is the
transition amplitude and $A_{2b}(i,S_2L_2 J_2)(s_i)$ describes
rescattering processes in the final two-particle channel (e.g. the
production of $\Delta(1232)$). In this spin-orbital momentum
decomposition we follow the formalism given in \cite{7,8,9}. The
exact form of the operators for the initial states
$Q^{in}_{\mu_1\ldots\mu_J}(S LJ)$ and final states
$Q^{fin}_{\mu_1\ldots\mu_J}(i,S_2L_2J_2S'L'J)$ can be found in
\cite{9}.
Following this decomposition, we use spectroscopic notation
$^{2S+1}L_{J}$ for the description of the initial state, the system of
two final particles and  the system "spectator  and two-particle
final state". For the initial $pp$ system, the states with total
momenta $J \leq 2$ and angular momenta $L=0,1,2,3$ between two
protons are taken into account. For the final three-particle
system, we restrict ourselves in the fitting procedure by angular
momenta $L_2=0,1,2$ and $L'=0,1,2$.
Due to nonresonant nature of the $pp$ system in the energy
region investigated here, there is no factorization of initial
and final vertices, and the transition amplitude depends on all
quantum numbers which characterize a partial wave (index $\alpha$).
Moreover, due to contribution of the triangle singularities, the
production parameters can be complex-valued. The best description
was obtained with the parameterization:
\begin{equation}
A^\alpha_{tr}(s)=\frac{a^\alpha_{1}+a^\alpha_{3}\sqrt{s}}
{s-a^\alpha_4}\, e^{ia^\alpha_{2}},
\label{trans}
\end{equation}
where $a^\alpha_i$'s are real values. The $a^\alpha_4$ parameter
corresponds to a pole situated in the region of left-hand side
singularities of the partial wave amplitudes. It is introduced to
suppress the growth of the amplitudes at large $s$.

We have also used another, more complicated parameterizations of the
transition amplitude. However, we obtained either worse description
of the data or similar one, with larger number of fitting parameters.
In the latter case, these results serve us to determine systematical
errors for  the various contributions to the cross sections.
For the $\pi N$ system in the intermediate state, we introduce
two resonances, $\Delta(1232)P_{33}$ and Roper $N(1440)P_{11}$. For
the $\Delta(1232)$, we use relativistic Breit-Wigner formula with
mass and width taken from PDG. The Roper state was
parameterized in agreement with Breit-Wigner couplings found in the
analysis \cite{our-roper}. Let us note that the present analysis is not
sensitive to the exact parameterization of the Roper resonance: only
the low energy tail of this state can influence the data.

For the description of the final $pp$ interaction we use a
modified scattering-length approximation formula:
\begin{equation}
A_{2b}^{\beta}(s_i)= \frac{\sqrt{s_i}}{1-\frac 12 r^\beta
q^{2}a^\beta_{pp}+ iqa^\beta_{pp}q^{2L}/F(q,r^\beta,L)},
\label{a_2b}
\end{equation}
where multiindex $\beta$ denotes possible combinations of
kinematical channel $i$ and quantum numbers  $S_2$, $L_2$ and $J_2$;
$a_{pp}^\beta$ is the $pp$-scattering length and $r^\beta$ is the
effective range of the $pp$ system. The $F(q,r,L)$ is the
Blatt-Weiss\-kopf form factor (it is equal to 1 for $L=0$ and the
explicit form for other partial waves can be found, for example, in
\cite{8}) and $q$ is the relative momentum in the final two-nucleon
system.
For the S-waves this formula corresponds exactly to the scattering
length approximation suggested in \cite{watson,migdal}. The $pp$
$^1S_0$ scattering length was fixed on value determined in the analysis
\cite{16}. In the analysis of the $pp\to pn\pi^+$ data the $pn$ scattering
length and effective range were fixed for S-waves at
$a(^1S_0)=-23.7$ fm, $r(^1S_0)=2.8$ fm and  $a(^3S_1)=5.3$ fm,
$r(^3S_1)=1.8$ fm.

\subsection{The PWA results and discussion}

We minimized the log-likelihood value fitting the present data on
the $pp\to pp\pi^0$ and $pp\to pn\pi^+$ reactions taken at the
proton momentum 1628 MeV/c together with obtained earlier data on
$pp\to pp\pi^0$ \cite{3,5,tubingen}. The data \cite{3,5} were taken
at PNPI and measured at nine energies covering the energy interval
from 600 up to 1000 MeV. The high statistics data at the momentum
950 MeV/c taken by the T\"ubingen group \cite{tubingen} were
included to fix the low energy region.

\begin{figure}[ht]
\centerline{\epsfig{file=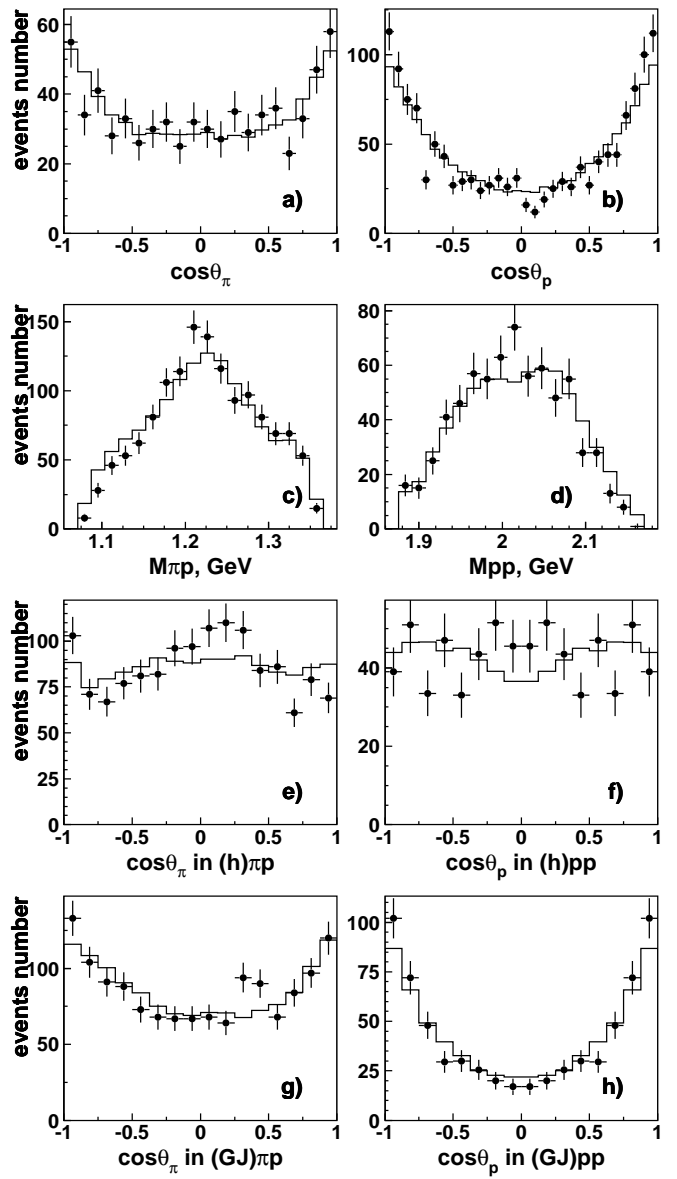,width=0.49\textwidth}}
\caption{\label{pi022ex}Angular distributions (a,b), effective-mass
spectra of final particles (c,d) in the reaction rest frame and
angular distributions of final particles in the helicity (e,f) and
Godfrey-Jackson (g,h) frames for the $pp\to pp\pi^0$ reaction taken
at the proton momentum 1628 MeV/c. The histograms show the result of
our partial wave analysis.}
\end{figure}

\begin{figure}[ht]
\centerline{\epsfig{file=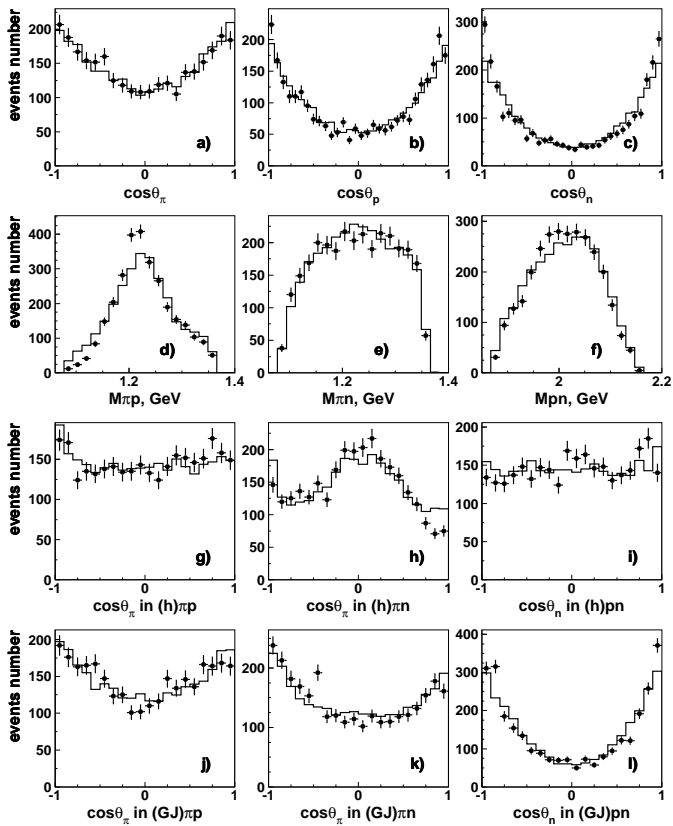,width=0.49\textwidth}}
\caption{\label{pip22ex}Angular distributions (a,b,c),
effective-mass spectra of final particles (d,e,f) in the reaction
rest frame and angular distributions of final particles in the
helicity (g,h,i) and Godfrey-Jackson (j,k,l) frames for the $pp\to
pn\pi^+$ reaction taken at the proton momentum 1628 MeV/c. The
histograms show the result of our partial wave analysis.}
\end{figure}

The experimental data (points with error bars) and the results of
the partial wave analysis (histograms) for the momentum 1628 MeV/c
are shown in Figs.~\ref{pi022ex} and \ref{pip22ex}. The first row
shows angular distributions of the final particles in the rest frame
of the reaction and the second row shows effective two-body mass
spectra. It is seen that our partial wave analysis describes these
distributions rather well.

The quality of the partial wave analysis is also demonstrated in
angular distributions in the helicity (third row) and
God\-frey-Jackson (fourth row) frames. These frames are the rest
frames of two final particle systems. In the helicity frame the
angle is calculated between one of the constituent particles and the
spectator particle. This frame is mostly suitable for the
investigation of cascade processes, when two colliding particles
form a system (e.g. resonance) which decays into a final two-body
system (e.g. another resonance) and a spectator. In the
Godfrey-Jackson frame the angle is calculated between one of the
constituent particles and the beam. This system is mostly suitable
to study production of the two-particle system due to the
$t$-channel exchange mechanism.

\begin{figure}[ht]
\centerline{\epsfig{file=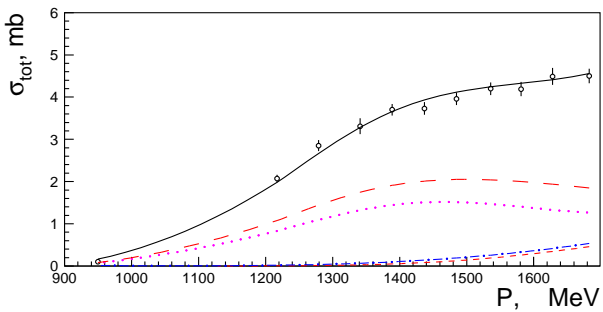,width=0.45\textwidth}}
\caption{\label{totpi0} The energy dependence of the total cross
section for the $pp\to pp\pi^0$ reaction. The experimental points
are taken from articles \cite{3,5,tubingen}. The solid curve is the
result of PWA; the long dashed curve shows the contribution from the
$^3P_2$ wave; the point curve from $^3P_1$; the point-dashed curve
from $^3P_0$ and short dashed curve from the $^1D_2$ wave.}
\end{figure}

The initial $^1S_0$ partial wave provides only a small contribution
to the $pp \rightarrow pp\pi^0$ reaction at the incident
proton momentum 1628 MeV/c. The largest contributions come from two
P-wave initial states: $^3P_2$ and $^3P_1$ (see Table~\ref{tab}). The
$^3P_0$ initial state contributes about 10\% to the total cross section
and we found notable contributions from the $^1D_2$ and
$^3F_2$ partial waves. The $^3F_2$ partial wave interferes rather
strongly with the strongest $^3P_2$ wave and its contribution is
defined with a rather large error.

All initial partial waves decay dominantly into the $\Delta(1232)p$
intermediate state. The contribution of channels with $\Delta(1232)$
production is varied for the different partial waves from 65 up to
90\%. The strongest non-resonant contribution is observed from the
$^3P_2$ initial state: here the transition $^3P_2\to(^3P_2)_{pp}\pi$
contributed in different fits from 20 to 35\% (from the contribution
of the $^3P_2$ partial wave). This
instability appears due to a notable interference in this channel
between the $(^3P_2)_{pp}\pi$ and $\Delta(1232)p$ intermediate states.

In the $pp \rightarrow pn\pi^+$ reaction at 1682 MeV/c, the contribution
of the $\Delta(1232)$ production to the cross section is even
stronger than in the case of the neutral pion production. The
non-resonant $pn$ partial waves with isospin 1 contribute much less to the
total cross section. For example, the transition
$^3P_2\to(^3P_2)_{pn}\pi$ was found to be less than 10\% from the contribution
of the $^3P_2$ initial state. However we observed a notable
transition $^1S_0\to(^3S_1)_{pn}\pi$ which is a dominant one for the
$^1S_0$ initial state and an appreciable contribution from the
$^3P_1\to(^1P_1)_{pn}\pi$ transition.

\begin{table}
\begin{center}
\caption{Contributions of the main partial waves to the single pion
production reactions at 1628 MeC/c}
\label{tab}       
\begin{tabular}{|cc|cc|}
\hline
&{\it pp $\rightarrow$ pp$\pi^0$} & &{\it pp $\rightarrow$ pn$\pi^+$} \\
\hline
$^1S_0$ & 1.0\er 0.5\,\%  & $^1S_0$ & 5.8\er 2.8\,\% \\
$^3P_0$ & 10.0\er 0.9\,\%  & $^3P_0$ & 11.5\er 0.9\,\% \\
$^3P_1$ & 26.0\er 7.7\,\%  & $^3P_1$ & 32.7\er 0.8\,\%  \\
$^3P_2$ & 44.0\er 1.6\,\%  & $^3P_2$ & 34.0\er 1.6\,\%  \\
$^1D_2$ & 8.0\er 1.1\,\%  & $^1D_2$ & 8.5\er 1.0\,\%   \\
$^3F_2$ & 11.4\er 7.7\,\%  &  $^3F_2$ & 6.3\er 1.5\,\% \\
\hline
\end{tabular}
\end{center}
\end{table}

Our partial wave analysis defines relative contributions of the
isovector waves to the total cross section of the single pion production
processes at the energy interval 400--1000 MeV.
Fig.~\ref{totpi0} shows the experimental behavior of the
$pp \rightarrow pp\pi^0$ cross section together
with the result of the partial wave analysis and
contributions of the dominant partial waves. It should be noted that
although we use the data on the $pp \rightarrow pn\pi^+$
reaction at 1628 MeV/c only the found partial waves predict
the total cross section at lower
energies in a good agreement with values given in
\cite{Shimizu:1982dx}).

\begin{figure}
\centerline{\epsfig{file=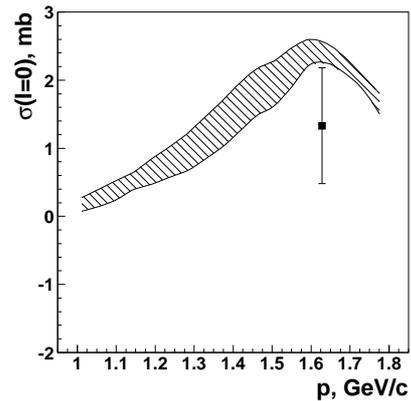,width=0.30\textwidth} }
\caption{ Isoscalar total cross section. The dashed band represents
results of work \cite{16}. The value calculated from
eq.(\ref{sig-i0}) using interpolation for $\sigma(np\to pp\pi^-)$
cross section from \cite{14,15} is shown as the black square with error bar.}
\label{fig_tcs}
\end{figure}

The present measurement of the cross section for the $pp \rightarrow
pp\pi^0$ reaction together with measurements of the cross section
for the $pn \rightarrow pp\pi^-$ reaction allows us to obtain the
isoscalar inelastic cross section by:
\begin{equation}
\label{sig-i0}
\sigma(I=0)=3[2\sigma(np\to pp\pi^-)-\sigma(pp\to pp\pi^0)].
\end{equation}

Figure~\ref{fig_tcs} shows the result of such calculation at 1628
MeV/c using the value for the $\sigma(np\to pp\pi^-)$ cross section
interpolated from the experimental data \cite{14,15} (black square).
The result of the work \cite{16} where the partial wave analysis of
the earlier $pp\to pp\pi^0$ data was performed together with the
$np\to pp\pi^-$ measurements taken with continues neutron beam is
shown by the band. The isoscalar cross section calculated using the
result of this analysis and new $pp\to pp\pi^0$ data at 1628 MeV is
fully consistent with this band and is not shown here.

\section{Conclusions}

   A detailed study of the differential cross section on the $pp \rightarrow
pp \pi^0$ and $pp \rightarrow pn \pi^+$ reactions has been performed
at the incident proton momentum 1628 MeV/c. The shape of the most
distributions is described qualitatively well by the OPE model,
although it fails to describe simultaneously the total cross
sections for the $pp \rightarrow pp\pi^0$ and $pp\rightarrow pn\pi^+$
reactions.

The partial wave analysis of the single pion production reactions
indeed reveals a dominant contribution from the $\Delta(1232)p$
intermediate state which explains a success of the OPE model.  But,
in addition, it allows us to obtain a combined description of all
analyzed reactions and extract contributions from the transition
amplitudes.

\section{Acknowledgements}

   We would like to express our gratitude to the bubble
chamber staff as well as to laboratory assistants, which toiled at
the film scanning and measuring.

\end{document}